\documentclass{aa}
\usepackage{txfonts}
\usepackage{epsfig}
\usepackage{subfigure}
\usepackage{multirow}

\usepackage{times}
\usepackage{natbib}
\bibpunct{(}{)}{;}{a}{}{,}

\begin{document}

\title{3C~33:~another case of photoionized soft X-ray emission in radio galaxies}

\author{E. Torresi\inst{1,2}, P. Grandi\inst{1}, M. Guainazzi\inst{3}, G.G.C. Palumbo\inst{2}, G. Ponti\inst{4}, S. Bianchi\inst{5}}

\offprints{Eleonora Torresi\\ \email{torresi@iasfbo.inaf.it}}

\institute{Istituto di Astrofisica Spaziale e Fisica Cosmica-Bologna, INAF, 
via Gobetti 101, I-40129 Bologna, Italy 
\and Dipartimento di Astronomia, Universit\`a di Bologna, via Ranzani 1, I-40127 Bologna, Italy 
\and European Space Astronomy Center of ESA, Apartado 50727, E-28080 Madrid, Spain 
\and APC Universit\'e Paris 7 Denis Diderot, 75205 Paris Cedex 13, France
\and Dipartimento di Fisica, Universit\`a degli Studi Roma Tre, via della Vasca Navale 84, 00146 
Roma, Italy} 

\date{Received / Accepted}

\authorrunning{E. Torresi et al.}

\abstract{All the observations available in the \emph{Chandra} and \emph{XMM-Newton} archives have been used to investigate the X-ray spectral properties of 3C~33. In this paper is presented a complete X-ray analysis of the nuclear emission of this narrow line radio galaxy. The broad band spectrum of 3C~33 is complex. The hard part resembles that of Seyfert 2 galaxies, with a heavily obscured nuclear continuum (N$_{H} \sim$10$^{23}$~cm$^{-2}$) and a prominent Fe K$\alpha$ line. This represents the nuclear radiation directly observed in transmission through a cold circumnuclear gas. On the other hand an unabsorbed continuum plus emission lines seem to fit well the soft part of the spectrum (0.5-2 keV) suggesting that the jet does not significantly contribute to the X-ray emission. We discuss the possible collisional or photoionized origin of the gas that emits the soft X-ray lines. Our results, strengthened by optical spectroscopy favor the photoionization scenario.


}

\titlerunning{3C33:another case of photoionized soft X-ray emission}

\maketitle

\section{Introduction}

With the advent of high resolution grating spectrometers, X-ray spectroscopy has become the most powerful tool for investigating the ionized component of the gaseous AGN environment. The detection of emission lines in Seyfert 2 galaxies has revealed gas photoionized by the nuclear engine (Sako et al. 2000; Kinkhabwala et al. 2002; Bianchi et al. 2005; Pounds \& Vaughan 2006; Guainazzi \& Bianchi 2007).\\
Spatially resolved studies realized with the \emph{Chandra} telescope on at least the three brightest AGNs, NGC 1068 (Kinkhabwala et al. 2002; Brinkman et al. 2002; Ogle et al. 2003), Circinus (Sambruna et al. 2001b) and Mkn 3 (Sako et al. 2000; Bianchi et al. 2005b; Pounds \& Page 2005) have revealed that their soft X-ray emission comes from regions extended on hundreds of pc, and the dimension and morphology of this emission closely resembles that of the Narrow Line Region (NLR) as mapped by the [OIII]$\lambda$5007 emission line (see e.g. Bianchi, Guainazzi \& Chiaberge 2006).\\
Systematic studies on large samples of  FRII Narrow Line Radio Galaxies (NLRG) show that the High Excitation Radio Galaxies (HEG; Jackson \& Rawlings 1997) have a heavily obscured continuum (intrinsic column density $\sim$10$^{23-24}$~cm$^{-2}$) related to the bright accretion disk obscured by an edge-on oriented thick torus, and a soft unabsorbed tail (Belsole et al. 2006; Hardcastle et al. 2006; Evans et al. 2006).\\
Soft X-ray emission above the extrapolation of the obscured component is often detected. The origin of the soft X-ray emission in  radio galaxies is still matter of debate. Radio-loud AGNs are potentially more complex than their radio-quiet counterparts, because the jet could also contribute to the X-ray emission.\\
It was suggested that the soft excess is related to the jet emerging from the edge of the dusty torus  (Hardcastle \& Worrall 1999; Belsole et al. 2006). Recently the discovery of emission lines produced by the photoionized gas in the soft X-ray spectrum of 3C~445 (Grandi et al. 2007; Sambruna et al. 2007) and 3C~234 (Piconcelli et al. 2008) has opened a new and alternative way to interpret the soft excess.\\
3C~33 is the right target to investigate the nature of the soft excess. It is a genuine type II radio galaxy, optically classified as High Excitation Radio Galaxy (HEG) and one of nearest HEG in the 3CRR catalogue (Laing Riley \& Longair 1983) located at \emph{z}=0.0597 \cite{popescu96}.\\ 
Its radio morphology is typical of FRII radio galaxies (L$_{178}$=7.5$\times$10$^{25}$~W~Hz$^{-1}$~sr$^{-1}$), showing simmetric lobes with hot spots on kpc scale (Leahy \& Perley 1991), and two simmetric jets in the VLBI image \cite{giova05}. The southern jet is stronger by a factor 2 at 5 mas. Expressing the jet/counterjet ratio (J) in terms of the bulk velocity of the jet ($\beta$) and the angle between the jet and the line of sight ($\theta$), \footnote{$\beta cos\theta=\frac{J^{1/p}-1}{J^{1/p}+1}$, \emph{p=2} \emph{$\gamma=2,10$}. From Urry \& Padovani (1995).} an inclination angle of~ $75^{\circ}-80^{\circ}$ can be deduced. Its spatial orientation is then favorable to reveal soft X--ray emission lines, because a strong de--amplification of the jet radiation is expected. \\
Moreover 3C~33 is very luminous in the [OIII] band (L$_{[OIII]}\sim$1.09$\times10^{42}$~erg s$^{-1}$) as it exhibits the highest [OIII] flux among the nearby NLRGs (z$<$0.3) of the 3CRR sample.
Previous nuclear observations of 3C~33 in the X-ray band go back to EINSTEIN observatory (Fabbiano et al. 1984) and ROSAT (Hardcastle Birkinshaw \& Worrall 1998; Hardcastle \& Worrall 1999) X--ray telescopes. EINSTEIN only provided an upper limit to the 0.5--3 keV luminosity, while ROSAT$/$HRI could detect the source in the soft X-ray band, although the HRI instrument on-board it could not resolve the source.\\
Evans et al. (2006) analyzed the \emph{XMM-Newton} EPIC PN data of 3C~33 showing for the first time the Seyfert 2 nature of the hard component. They modeled the soft component with an unabsorbed power law.\\
Kraft et al. (2007), taking advantage of the unprecedented imaging of \emph{Chandra}, could perform a very detailed study of 3C~33. They could resolve the north and south hot spots regions. They also studied the nuclear component with a basic model consisting of an absorbed power law plus a Fe K$\alpha$ line and a thermal component. Large residuals between 1 and 3 keV could be explained either by assuming the presence of a reflection component or of a less absorbed power law, that they interpreted as jet emission on parsec scale. Kraft et al. (2007) used thermal gas to model the soft X-ray emission on the base of the \emph{Chandra} imaging that reveals in the \emph{soft} band (0.3-0.7 keV) a clearly extended structure around the unresolved hard nucleus. A similar elongation was observed in the WFPC1 and WFPC2 Hubble Space Telescope imaging, and was interpreted by the authors as a tracer of stellar distribution.\\ 
In this work we  reanalyzed all the \emph{Chandra} and \emph{XMM-Newton} observations of the radio galaxy 3C~33 available in the public archives, and propose a different interpretation of the nature of the soft X-ray emission of 3C~33. \\

\section{Observations and data reduction}

\emph{Chandra} ACIS-S observed 3C~33 twice [$\sim$20 ks on 2005 November 8 (ObsID 6190) and 2005 November 12 (ObsID 7200)]. Spectra and instrument responses were generated using the Chandra Interactive Analysis of Observations (CIAO) 3.4 and the Chandra Calibration Database (CALDB) 3.4.3. The source spectrum was extracted from a circular region with radius 1.5$''$ centered at the peak of the X-ray emission, while background counts were extracted from a 4$''$ radius circle in the same CCD, but outside of the extended emission of the source.
 The \emph{Chandra} count rate from the nucleus in both observations is $\sim$0.06 cts s$^{-1}$, implying a negligible pileup effect.\\
In the hard band (2--8 keV) the X-ray morphology of the 3C~33 nucleus is point--like. An extended structure around the nucleus becomes evident in the soft band (0.5--2 keV), looking approximately elliptical, with a major axis of $\sim5''$, elongated in the northeast-southwest (NE-SW) axis, in agreement with the analysis of Kraft et al. (2007).\\
\emph{XMM-Newton} observed 3C~33 twice [on 2004 January 4 and January 21 (ObsID 0203280101; 0203280301)]. We checked both observations but we analyzed only the second pointing ($\sim$ 9 ks)  because the first one is affected by high flaring activity. This prevented us from obtaining useful information from the RGS analysis.\\
The EPIC cameras were operated in `prime full window' mode. The data were reduced using the SAS (v. 7.1) with standard procedures.\\
The light curve over 10 keV was extracted to check high background periods. We excluded the time intervals characterized by count rate higher than 1.0 cts/s in the EPIC-PN, and 0.35 cts/s in the EPIC-MOS. After this data cleaning we obtained a net exposure of 6.3 ks for the PN and 8.5 ks for the two MOS cameras.\\
The source and the background spectra were extracted from circular regions of 35$''$ radius. The background was chosen in the same CCD of 3C~33, but in a region not contaminated by the source. The source counts are $\sim$1500 and $\sim$1000 in the PN and the two combined MOS, respectively.
The response matrices were created using the SAS commands RMFGEN and ARFGEN.\\
Events outside the 0.5-10 keV band were discarded in both PN and  MOS spectra (Kirsch 2006).\\
All spectral fits presented in this paper include absorption due to a line-of-sight Galactic column density of N$_{H}$=3.06 $\times10^{20}$cm$^{-2}$ (Kalberla et al. 2005).\\
Errors are quoted at the 90$\%$  confidence level ($\Delta \chi^{2}$=2.7) for one interesting parameter.\\ 
The cosmological parameters used throughout the paper are  $H_0=70$ km$^{-1}$ s$^{-1}$ Mpc$^{-1}$, $\Omega_{m}=0.3$, $\Omega_{\Lambda} = 0.7$ (Spergel et al. 2007). For 3C~33, z=0.0597 corresponds to a luminosity distance of 263.9 Mpc, and 1$''$ corresponds to a physical scale of 1.14 kpc.\\
The spectral analysis obtained by both telescopes were performed using the XSPEC 11.3 package.\\

\section{Spectral analysis}

We started our study using \emph{Chandra} data.  The two observations were simultaneously analyzed using the spectra extracted from 1.5$''$ regions around the nucleus. We also considered a larger extraction radius of the nuclear source (r=4.5$''$) in order to include the extended soft X--ray emission. As expected the statistical quality of the spectrum decreased  because of the higher background. Indeed, very few are the source counts in the (0.5--2) keV band from an elliptical region of 5.5$''$ semi--major axis and 4$''$ semi--minor axis (excluding the nucleus), less than 3$\%$ of the total counts in the 1.5$''$ region. In physical terms most of the soft--X radiation comes from $\sim 2$ kpc around the source. We then decided to consider the smaller nuclear region in order to maximize the spectrum signal to noise ratio (S$/$N).\\

\subsection{Hard X--ray spectrum}
The broad band spectrum appeared quite complex. We initially decided to fit only the energy range between 3-10 keV in order to focus on the hard X-ray continuum excluding the soft emission. 
At first we applied a photoelectrically absorbed power-law plus a Gaussian line profile. The restricted energy band and the limited statistics do not allow to constrain very well the spectral index (the column density and the spectral index being strongly correlated), so we decided to fix $\Gamma$=1.7 following Kraft et al. (2007). This choice was confirmed to be the right one by the \emph{XMM-Newton} analysis (see below). As the $\chi^{2}$=102(80) was not completely satisfactory we tested  for the presence of a reflection component, as in obscured AGNs is often required [see for example, Mkn3 (Pounds \& Page 2005), and 3C445 (Grandi et al. 2007; Sambruna et al. 2007)]. When a (\emph{PEXRAV}) \cite{pexrav} model was applied the $\chi^{2}$ decreased to 82 for 79 d.o.f., giving a better representation of the data (P$_{F}$ 99.99$\%$) . The reflection (R=$\Omega/2 \pi$) was let free to vary, while the high-energy cut-off of the incident power law was set at 160 keV (Grandi, Malaguti \& Fiocchi 2006), and the elemental abundances were assumed to be solar.
We repeated the same excercise leaving $\Gamma$ free to vary. Also in this case the addition of a reflection component improves the fit, even if the strong correlation between the reflection component, $\Gamma$ and N$_{H}$ prevent us to substantially constrain these three variables.\\
The FeK${\alpha}$ emission line is not resolved. The rest--frame energy is  E$_{K\alpha}$=6.38$^{+0.05}_{-0.06}$ keV, the upper limit of the intrinsic width is $\sigma<$ 180 eV and the equivalent width is EW$_{K\alpha}$=172$^{+84}_{-76}$ eV.\\ 

\subsection{Soft X--ray spectrum}

Once a good fit of the hard component was obtained we extended our study to the soft part (0.5--3 keV) of the spectrum. The simple extrapolation of the hard model introduced a strong soft component. 
Currently, in the literature, at least three different interpretations of the soft excess in obscured radio loud AGNs are proposed: i) thermal radiation from collisional gas (Kraft et al. 2007; Evans et al. 2006); ii) non--thermal emission from a jet (Belsole et al. 2006); iii) nuclear scattered radiation plus emission lines produced by a circumnuclear photoionized gas (Bianchi \& Guainazzi 2007; Grandi et al. 2007; Sambruna et al. 2007; Piconcelli et al. 2008).  All these three possibilities were tested.\\
At first we tried to reproduce it with a thermal model (\emph{mekal}) absorbed by the Galactic column density. This implicitly assumes that the soft excess is produced by a collisional plasma (with solar abundances). The gas temperature required by the fit is $kT\sim 0.7$, and the model partially reproduced the soft excess, $\chi^{2}$(d.o.f.)=138(104), but the inspection of the residuals still showed excesses below 2 keV. A second \emph{mekal} was added to verify whether a multi-temperature gas is present, but the second temperature shifted to a very improbable value larger than 10 keV, and excesses were still present in the soft X-ray band (see Fig.1 \emph{left panel}). Note that if the abundances  are free to vary, they become extremely low, miming a power-law. Indeed  replacing the ``second'' thermal emission model with a simple power-law ($\Gamma_{s}$ free to vary), that represents a non-thermal emission, the $\chi^{2}$ improved $\chi^{2}$(d.o.f.)=124(103), and the thermal emission is no longer required
anymore (the mekal model is consistent with zero). Moreover, although this was the best fit obtained so far, the model could not completely reproduce the soft X-ray data (Fig.1 \emph{right panel}).\\
Finally the third hypothesis was verified. A soft excess was fitted with a power law plus lines, as expected in case of circumnuclear photoionized gas. The soft power law slope ($\Gamma_{s}$) was linked to the hard one ($\Gamma_{h}$), in order to represent the scattered nuclear continnum.
 The fit (see Fig.\ref{fig2} \emph{left panel}) was very satisfactory $\chi^{2}$(d.o.f.)=103(99). A summary of the best-fitting parameters for the hard X-ray spectra is shown in Tab.\ref{tab1}, while the detected lines and relative significance levels are listed in Tab.\ref{tab2}.\\

In order to find indipendent confirmation of this result, we explored the \emph{XMM-Newton} EPIC spectra.  The PN and MOS results were reassuring. All the EPIC spectra (see Fig.\ref{fig2} \emph{right panel}) were very well fitted with the same best-fit model of the \emph{Chandra} continuum. Interesting enough, the PN data needed soft lines at the same energies measured by \emph{Chandra}. \\
The PN best-fit parameters are listed in Tab. \ref{tab3}. The relative $\chi^{2}$ is 37.5 for 37 degrees of freedom.\\

\section{Results}

Our analysis displays typical features produced by a cold circumnuclear gas: an unresolved ($\sigma <$ 180 eV) narrow iron line (EW $\simeq$ 200 eV), and a reflection component (R $\simeq$ 2) (Risaliti 2002), showing the Seyfert 2 nature of the hard spectrum of 3C~33.
Both these values are consistent with the \emph{Chandra} results of Kraft et al. 2007 (their Model 1).\\
The reflection component was obtained assuming an arbitrarily inclination angle of $\theta\simeq$ 60$^{\circ}$ (PEXRAV parameter cos$\theta\simeq$ 0.45). However even considering a larger inclination angle, as suggested by the radio data, the value of R does not significantly change. This is not surprising as the limited statistics and the large number of parameters required by the fit. Although the reflection component is slightly higher than that expected from an homogeneous dusty torus, it is not unusual and often observed in radio--quiet obscured objects. 
This probably indicates the presence of a more complex inhomogeneous and structured circumnuclear absorber (Risaliti et al. 2002).\\

The spectral analogy between Seyfert 2s and obscured radio galaxies seems to be maintained at soft X--ray energies, where the best fit requires emission lines overimposed on a weak continuum. 
The lines in Tab. \ref{tab2} and \ref{tab4} are likely produced by H-- and He--like ions of light elements such as Oxygen and Neon, often detected in obscured AGNs showing strong evidence of photoionized gas (Bianchi \& Guainazzi 2007; Grandi et al. 2007; Sambruna et al. 2007; Piconcelli et al. 2008). 
Although the moderate ACIS energy resolution does not allow to draw strong conclusions, we stress that the centroid of the lines occurs where two strong features produced by photoionized gas are expected, i.e. the OVII recombination feature and the NeIX triplet lines. Incidentally we note that  the intrinsic nuclear luminosity of 3C~33 (L$_{1-1000ryd} \sim$2$\times$10$^{44}$ erg sec$^{-1}$) is  comparable to those observed in typical Seyfert 2s, for which the higher quality of the data has permitted detailed spectral studies of the soft X-ray diagnostics (see, for example, Pounds \& Page 2005, Sako et al. 2000ab). 
 Approximate computations indicate that the energy budget of the AGN is sufficient to produce a range of values of the ionization parameter ($\xi=\frac{L_{1-1000ryd}}{n_{e}R^{2}}$) consistent with the photoionized scenario (Kinkhabwala et al. 2002).
If the soft X--ray emission is related to the Narrow Line Region (as in Seyfert 2s), assuming a minimum  (R$_{min} \sim$1 pc) and a maximum (R$_{max} \sim$2 kpc, our nuclear extraction region) distance of the gas, a typical NLR electron density  n$_{e}\sim$ 10$^{3}$ cm$^{-3}$ implies log$\xi$ spanning between -2 and 4 erg cm s $^{-1}$.


\begin{table}
\caption{\emph{Chandra} ACIS best-fit parameters. Normalizations $n$ are expressed in cm$^{-2}$~s$^{-1}$~keV$^{-1}$; the normalization of the iron line is expressed in photons cm$^{-2}$~s$^{-1}$.}
 \label{tab1}      
\centering          
\begin{tabular}{l c }     
\hline\hline       
$\Gamma$             &   1.7$^{*}$  \\
R                    &   2.18$^{+0.48}_{-0.60}$      \\
N$_{G}$            & (3.06$\times10^{20}$) cm$^{-2}$ \\
N$_{H}$            & (5.28$^{+1.04}_{-0.42}) \times10^{23}$ cm$^{-2}$ \\
n                    & (1.83$^{+0.51}_{-0.33}) \times10^{-3}$ \\
n$_{unobscured}$     & (4.48$^{+3.43}_{-2.11}) \times10^{-6}$\\ 
$E_{Fe~~}$  (keV)   & 6.38$^{+0.05}_{-0.06}$  \\ 
$F_{Fe~~}$           & (2.49$^{+1.21}_{-1.11})\times10^{-5}$\\
EW (eV)              & 172$^{+84}_{-76}$ \\ 
\hline
\bf{$\chi^{2}$(d.o.f)}     &\bf{103(99)}\\
\hline\hline
\multicolumn{2}{l}{Unabsorbed luminosity and absorbed flux} \\
$L_{2-10~~keV}$ (erg~s$^{-1}$)      &     7.31 $ \times10^{43}$ \\
$Flux_{2-10~~keV}$ (erg~s$^{-1}$ cm$^{-2}$)&  2.67 $\times 10^{-12}$ \\
\hline\\                                
\multicolumn{2}{l}{(*) value fixed}\\
\end{tabular}
\end{table}



\begin{table}
\caption{Emission lines detected in the \emph{Chandra} ACIS spectrum. $E_c$ is the centroid line energy (keV) in the source rest frame; $F$ is the line flux in units of 10$^{-6}$ photons cm$^{-2}$  s$^{-1}$; $P_F$ is the significance of the fit improvement after including the Gaussian line in the model, based on the F-test.}
\label{tab2}      
\centering          
\begin{tabular}{l c c c c}     
\hline\hline\    
$Energy$                 & $Intensity$            &$P_{F}$     & $Tentative~IDs$\\
\hline                
0.76$^{+0.02}_{-0.13}$    &2.93$^{+1.16}_{-1.02}$  &$>98.9 \%$ &OVII RRC, Fe XVII 3s-2p\\
0.93$^{+0.04}_{-0.02}$   &2.51$^{+0.85}_{-1.20}$  &$>99 \%$ &Ne IX rif\\                      
1.30$^{+0.04}_{-0.02}$   &1.41$\pm0.46$           &$>99.97 \%$ &Mg XI He$\alpha$\\
\hline \\ 
\end{tabular}
\end{table}



\begin{table}
\caption{\emph{XMM-Newton} EPIC best-fit parameters. Normalizations $n$ are expressed in cm$^{-2}$~s$^{-1}$~keV$^{-1}$; the normalization of the iron line is expressed in photons cm$^{-2}$~s$^{-1}$. }
 \label{tab3}      
\centering          
\begin{tabular}{l c }     
\hline\hline       
$\Gamma$             &   1.69$^{+0.18}_{-0.10}$  \\
R                    &   2.7$^{+0.81}_{-0.60}$      \\
N$_{G}$            & (3.06$\times10^{20}$) cm$^{-2}$ \\
N$_{H}$            & (7.14$^{+1.75}_{-1.21}) \times10^{23}$ cm$^{-2}$ \\
n                    & (1.83$^{+0.47}_{-0.31}) \times10^{-3}$ \\
n$_{unobscured}$     & (8.1$^{+4.08}_{-7.28}) \times10^{-6}$\\ 
$E_{Fe~~}$   (keV)   & 6.40$^{+0.13}_{-0.08}$  \\ 
$F_{Fe~~}$           & (3.65$^{+3.93}_{-2.04})\times10^{-5}$\\
EW (eV)              & 206$^{+199}_{-115}$ \\ 
\hline
\bf{$\chi^{2}$(d.o.f)}  &\bf{37.5(37)}\\
\hline\hline
\multicolumn{2}{l}{Unabsorbed luminosity and absorbed flux} \\
$L_{2-10~~keV}$ (erg~s$^{-1}$)      &     7.53$ \times10^{43}$ \\
$Flux_{2-10~~keV}$ (erg~s$^{-1}$ cm$^{-2}$)&  2.64 $\times 10^{-12}$ \\
\hline\\   
\end{tabular}
\end{table}



\begin{table}
\caption{Emission lines detected in the \emph{XMM-Newton} EPIC spectrum. $E_c$ is the centroid line energy (keV) in the source rest frame; $F$ is the line flux in units of 10$^{-6}$ photons cm$^{-2}$  s$^{-1}$; $P_F$ is the significance of the fit improvement after including the Gaussian line in the model, based on the F-test.}
\label{tab4}      
\centering          
\begin{tabular}{l c c c c}     
\hline\hline\    
$Energy$                 & $Intensity$            &$P_{F}$     & $Tentative~IDs$\\
\hline                
0.71$\pm0.03$            &7.89$^{+2.8}_{-4.6}$    &$>99.8 \%$  &OVII RRC, Fe XVII 3s-2p\\
0.93$^{+0.03}_{-0.05}$   &5.23$^{+3.1}_{-4.8}$    &$>99 \%$    &Ne IX rif\\                        
1.26$^{+0.04}_{-0.17}$   &2.35$\pm2.2$            &$>88 \%$    &Mg XI He$\alpha$\\
\hline \\ 
\end{tabular}
\end{table}


\begin{figure}
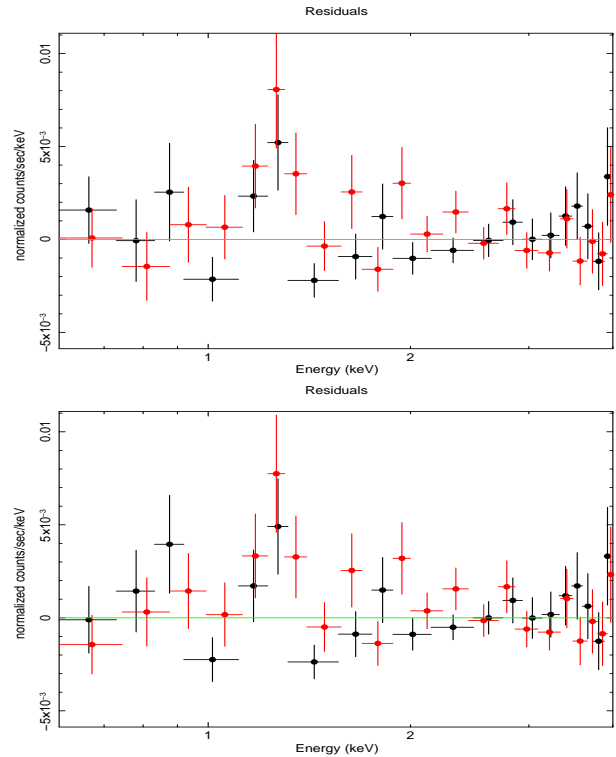

\epsfig{file=2mekal_res_paper.cps,height=8cm,width=5cm, angle=-90}
\epsfig{file=powerlaw_res_paper.cps,height=8cm,width=5cm, angle=-90}
    \caption{Residuals corresponding to the two different soft excess models: \emph{left} two mekal and \emph{right} a scattered power--law.}
\label{fig1}
\end{figure}

\begin{figure}
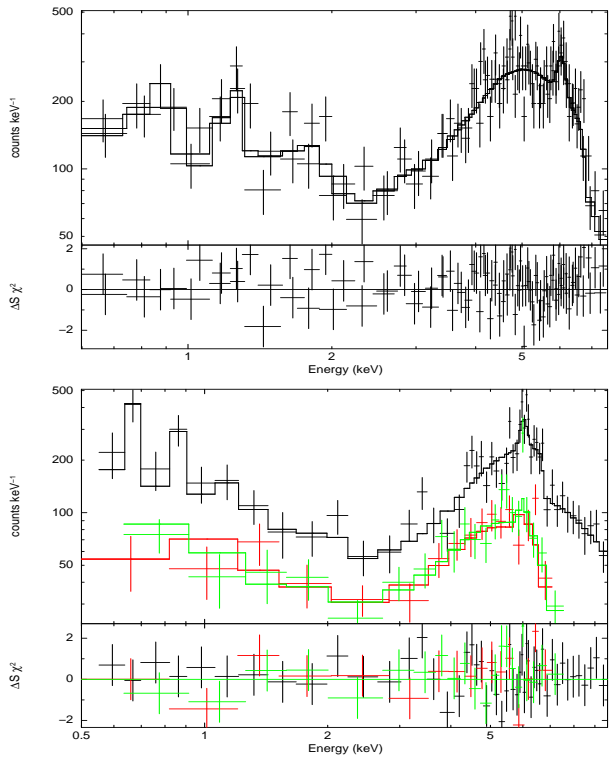

\epsfig{file=chandra.ps,height=8cm,width=5cm, angle=-90}
\epsfig{file=pn_mos1_2.cps,height=8cm,width=5cm, angle=-90}
     \caption{\emph{Chandra} spectra (\emph{left panel}) and \emph{XMM} spectra (\emph{right panel}) of 3C~33.  PN, MOS 1 and MOS 2 spectra are represented with \emph{black}, \emph{red} and \emph{green} crosses, respectively. Residuals are in terms of standard deviations of the observed data from the model.}
\label{fig2}
\end{figure}

\section{The origin of the soft X-ray emission}

\subsection{Imaging} 

\begin{figure}
\center\epsfig{file=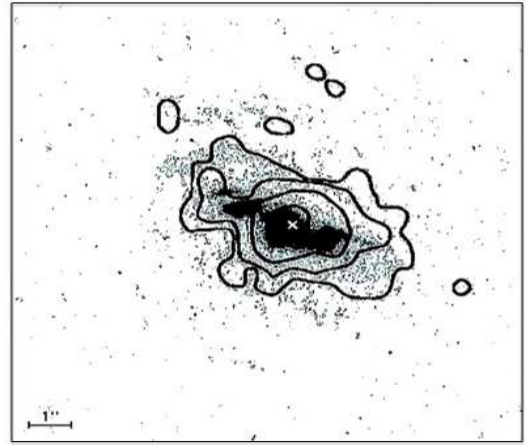,height=6cm,width=7cm, angle=0}
     \caption{Superposition of the \emph{Chandra} soft X-ray ($<$2 keV) contours on an \emph{HST} image taken through a linear ramp filter at redshifted [OIII]$\lambda$5007. The sign ``x'' indicates the centre of the hard X-ray source, north is up, east to the left. The X--ray image was smoothed with a Gaussian of FWHM $\sim$ 6 pixels. The contours correspond to four logarithmic intervals in the range 1-60$\%$ of the peak flux.}
\label{fig3}
\end{figure}

X-ray spectroscopy provides evidence for the presence of soft X--ray emission lines in the radio galaxy 3C~33.\\ 
In spite of the limited spectral power of the CCDs we could detect features that are plausibly emitted by a photoionized gas (see Tab. \ref{tab2} and Tab. \ref{tab4}).\\
The analogy with Seyfert 2 galaxies is straightforward. Narrow Line Radio Galaxies would share similar continua and environments with their radio-quiet counterparts. 
In Seyfert 2 galaxies the coincidence in morphology and dimension of the soft X-ray and the [OIII]$\lambda$5007 emission suggests the Narrow Emission Line Region as the location of such gas. It is then natural to search for a similar X-ray and optical spatial correlation in 3C~33.\\
For this reason the narrow--band HST ACS/WFC1 optical image, centered at redshifted [OIII]$\lambda$5007 (filter FR551N), was retrieved from Multi Mission Archive at STScI (MAST) and was aligned with the \emph{Chandra} image as described in Bianchi, Guainazzi \& Chiaberge (2006).\\ 
Fig. \ref{fig3} shows the \emph{Chandra} soft X-ray contours ($<$ 2 keV) superimposed on the \emph{HST} [OIII]. The coincidence in extension and morphology between the two emissions is impressive.

\subsection{Further hints from optical spectroscopy}

Further support in favor of the photoionization scenario comes from optical spectroscopy.\\
Several line diagnostic diagrams  have been proposed to discriminate between collisionally ionized plasma (e.g. starburst emission) and photoionized plasma irradiated by a central engine (see for example Miller et al. 2003). One of these, based on the  [OIII]$/$H$\beta$ and [NII]$/$H$\alpha$ ratios (Fig. 3 of Miller et al. 2003) could be used to test the nature of the gas emission of 3C~33. Using the nuclear reddening-corrected emission line luminosities reported by Koski (1978) we estimated  log([OIII]$/$H$\beta$)$\sim$~1.1 and log([NII]$/$H$\alpha$)$\sim$~-0.17. Such values place 3C~33 in the region populated by the photoionized AGNs.\\ 
Baum et al. (1992) performed an accurate study of the extended kpc emission-line region (EELR). They also found a large value of [OIII]$/$H$\beta$ ratio that they interpreted as produced by photoionized gas with a high ionization parameter.\\
This result strengthens the early conclusion of  Baum \& Heckman (1989) who demonstrated that the number of photons needed to photoionize the emission-line nebula of this source equals the number of ionizing photons coming from the central engine.\\
In addition, on the base of the kinematic classification of powerful radio galaxies proposed by Baum et al. (1992),  3C~33 belongs to the \emph{rotator} class. Rotators, characterized by  a systematic gradient in the velocity of the emission-line gas, are nearly always associated with powerful radio sources (L$_{radio}\geq$10$^{42}$~erg~s$^{-1}$) having luminous high-ionization nuclear emission-line region.\\
The kinematic study of the gas also shows that the gas rotation axis and the radio axis are aligned within 10$^{\circ}$ (Baum et al. 1992). Moreover the rotation axis of the gas is not aligned with the stellar rotation axis (Heckman et al. 1985), disfavoring the hypothesis of a collisional (starburst) gas emission.

\section{Conclusions}

An X-ray analysis of the Narrow Line Radio Galaxy 3C~33 is presented. All the \emph{Chandra} and \emph{XMM-Newton} observations available in the public archives are used. Our study confirms the Seyfert 2 nature of this source. In agreement with previous works we confirm that the broad-band spectrum is complex and largely dominated by strong obscuration (N$_{H}\sim$10$^{23}$~cm$^{-2}$).\\ 
A prominent Fe K$\alpha$ line at 6.4 keV is also present. Its EW of $\sim$ 200 eV is in agreement with being produced by optically-thick circumnuclear matter.\\
We advocate photoionized gas for an interpretation of the soft X-ray emission.
The clues favoring this hypothesis are:

\begin{enumerate}
\item both \emph{Chandra} and \emph{XMM-Newton} soft X-ray spectra can be better fit with highly ionized lines and unobscured scattered power-law, rather than with thermal components;
\item a rough estimate of the ionization parameter $\xi$, expected in the case that the emitting line gas is located in the NLR, is consistent with a photoionized scenario;  
\item there is a remarkable spatial coincidence between the extended soft X-ray  and the [OIII]$\lambda$5007 emission;\\

X-ray findings are also consistent with pieces of evidence coming from other wavelengths such as:\\

\item the nuclear line ratios [OIII]$/$H$\beta$ and [NII]$/$H$\alpha$ put 3C~33 in the right position of the line diagnostic plane of Miller et al. (2003). It is among the AGNs, far away from the region populated by starburst galaxies;
\item the extended kpc emission region has a line rich optical spectrum with large [OIII]$/H\beta$ ratio, typical of photoionized gas;
\item the estimated photons required to photoionize the emission-line nebula (in the optical band) are provided entirely by the nuclear engine;
\item the kinematical properties of the extended gas of 3C~33, based on optical spectroscopy, are typical of powerful radio galaxies with high-ionization nuclear emission;
\item  the gas rotation axis and the radio axis are aligned within 10$^{\circ}$, but the former is not aligned with the stellar rotation axis. 
\end{enumerate}

\section*{Acknowledgments}

We thank very much Carlo Stanghellini, Edoardo Trussoni for helpful discussions, and Marco Mignoli for providing us useful references. We thank the referee Rita Sambruna for valuable suggestions and comments. ET thanks for support the Italian Space Agency (contract ASI-I/088/06/0). GP acknowledges ANR and ASI-INAF for support (contracts ANR-06-JC-0047, ASI-INAF I/023/05/0 and ASI-I/088/06/0). SB acknowledges financial support from ASI-I/088/06/0.

{}

\end{document}